\documentclass[prb,aps,showpacs,twocolumn]{revtex4}
\usepackage{amsmath,color}
\usepackage{graphicx}
\usepackage{epstopdf}

\begin{document}
\title{Multipair DC-Josephson Resonances in a biased all-superconducting Bijunction}
\date{\today}

\author{T. Jonckheere$^{1,2}$, J. Rech$^{1,2}$, T. Martin$^{1,2}$,
  B. Dou\c{c}ot$^3$, D. Feinberg$^4$, R. M\'elin$^4$} 
\affiliation{$^1$ Centre de Physique Th\'eorique, Aix-Marseille Universit\'e, CNRS, CPT, UMR 7332, 13288 Marseille, France}
\affiliation{$^2$ Universit\'e de Toulon, CNRS, CPT, UMR 7332, 83957 La Garde, France}
\affiliation{$^3$ Laboratoire de Physique Th\'eorique et des Hautes Energies,
  CNRS UMR 7589, Universit\'es Paris 6 et 7, 4 Place Jussieu, 75252 Paris
  Cedex 05}
\affiliation{$^4$ Institut NEEL, CNRS
  and Universit\'e Joseph Fourier, UPR 2940, BP 166, F-38042 Grenoble Cedex 9, France}


\begin{abstract}
An all-superconducting bijunction consists of a central superconductor contacted to two lateral
superconductors, such that non-local crossed Andreev reflection is operating. Then new correlated transport channels for the Cooper pairs
appear in addition to those of separated conventional Josephson junctions. We study this system in a configuration where
the superconductors are connected through gate-controllable quantum dots.
Multipair phase-coherent resonances and phase-dependent multiple Andreev reflections are both obtained when the voltages of the lateral superconductors are commensurate,
and they add to the usual local dissipative transport due to quasiparticles. The  two-pair resonance (quartets) as well as some other higher order
multipair resonances are $\pi$-shifted at low voltage. Dot control can be used to dramatically enhance the multipair current when the voltages are resonant with the dot levels.  
\end{abstract}

\maketitle

\section{Introduction}
\label{sec:introduction}
One of the most striking manifestation of macroscopic quantum coherence is the Josephson effect
\cite{TINKHAM}: a DC current flows when a phase difference is imposed on a junction bridging two superconductors with a narrow insulating, metallic or semiconducting region.
When applying a constant voltage bias to this same junction, an oscillatory current arises \cite{JOSEPHSON} and the application of an rf-irradiation leads to the observation of Shapiro steps with zero differential resistance \cite{JOSEPHSON,SHAPIRO} and phase coherence \cite{SQUIDSHAPIRO}.
More generally, the microscopic origin of these effects is Andreev reflections of electrons and holes
at the boundaries of the two superconductors. The same mechanism participates in the appearance of a subgap structure in highly transparent voltage-biased junctions, a feature understood to be due to dissipative quasiparticle emissions called 
multiple Andreev reflections (MAR) \cite{MAR,CUEVAS}, which were
observed in atomic point contact experiments \cite{SACLAY}.

Non-local quantum mechanical phenomena \cite{EPR} and entanglement are nowadays investigated in condensed matter physics, in particular in superconducting circuits \cite{entanglement_exp}. Multiterminal superconducting hybrid devices with one superconducting arm and two normal metal 
electrodes have also been studied in the last decade \cite{car theory}  with the aim of detecting non-local
entangled electron pairs \cite{entanglement_nano}. There is now convincing experimental data on non-local 
current and noise detection which points in this direction \cite{car experiment,TAKIS,BASEL}. 
Yet, there is also a growing interest in three-terminal all-superconducting hybrid structures \cite{HOUZET-DUHOT,HOUZET-SAMUELSSON,VINOKUR,LEFLOCH-NOISE}, so far mainly in regimes dominated by 
phase-insensitive processes. A recent calculation for a SNS junction, where the N region is tunnel-coupled to another superconductor, also showed resonances ascribed 
to voltage-induced Shapiro steps \cite{CUEVAS-POTHIER}. 

The present work shows that a non-dissipative phase-coherent Josephson signal of Cooper pair transport could be observed in a device consisting of three superconductors driven out of equilibrium.
This effect relies on a combination of both direct Andreev reflections and non-local crossed Andreev reflections (CAR) \cite{car theory}, and is thus directly tied to non-local entangled electron processes as well as Josephson physics. 
Here, the ``bijunction'' which we propose consists of a central superconductor $S_0$ 
coupled via two adjustable quantum dots to two lateral superconductors $S_a$ and $S_b$, biased
at voltages $V_a$ and $V_b$ (Fig.~\ref{fig:bijunction}). 
As the coherence length of $S_0$ (which is grounded at $V_0\equiv 0$) is assumed to be larger than the distance between the dots, 
this bijunction cannot be simply considered as two separated junctions in parallel. 
Each junction consists of a quantum dot, made with e.g. carbon
nanotubes \cite{NANOSQUID,TAKIS} or nanowires \cite{BASEL}, and labeled $D_\alpha$ ($\alpha = a,b$). The dots introduce additional degrees of freedom (position of energy levels, coupling widths) which provide full control of the junctions.
Equilibrium calculations \cite{FREYN} in a similar three-terminal device involving normal-metal interfaces showed that a bijunction 
could be a source of spatially correlated pairs of Cooper pairs (referred to as "non-local quartets")
transmitted into $S_a$ and $S_b$ simultaneously.

This article reports on calculations of out-of-equilibrium transport in a 
biased $S_a D_a S_0 D_b S_b$ bijunction, with as main results:  

(i) At commensurate voltages $n V_a + m V_b =0$ ($m$ and $n$ integers),
{\it DC Josephson resonances} appear, 
which correspond to the phase-coherent transport of $n$ pairs to $S_a$, and $m$ pairs to $S_b$, from $S_0$. 

(ii) The Josephson current-phase relation
of quartet resonances ($n=m=1$) and that of some higher-order resonances are
$\pi$-shifted at low bias. This new mechanism for producing  a $\pi$-shift is of
particular importance for future interferometry experiments. 

(iii) Gate and/or bias voltages can be tuned to enhance the
multipair resonances by orders of magnitude as compared to the adiabatic
regime, making them easily observable in experiments. 

(iv) At larger biases, a {\it DC} quasiparticle-pair interference term, corresponding to phase-dependent MAR, emerges from the 
dissipative Josephson component.

The structure of this article is the following. In Sec.~\ref{sec:adiab}, we explain qualitatively the multi-pair Josephson resonances from a simple adiabatic argument. The following sections are concerned with an exact out-of-equilibrium
calculation, valid at arbitrary voltages. Sec.~\ref{sec:hamil} details the Hamiltonian formalism which we have used to 
perform the calculations. Sec.~\ref{sec:metallic} shows and discusses results obtained in the regime where the quantum dots
 have a behavior similar to metallic junctions. The next section shows results for the opposite regime where the dots
 present a narrow resonance. Finally, Sec.~\ref{sec:conclusions} presents the conclusions and perspectives of this work.

\section{Adiabatic argument}
\label{sec:adiab}
\begin{figure}[tb]
\includegraphics[width=0.49\columnwidth]{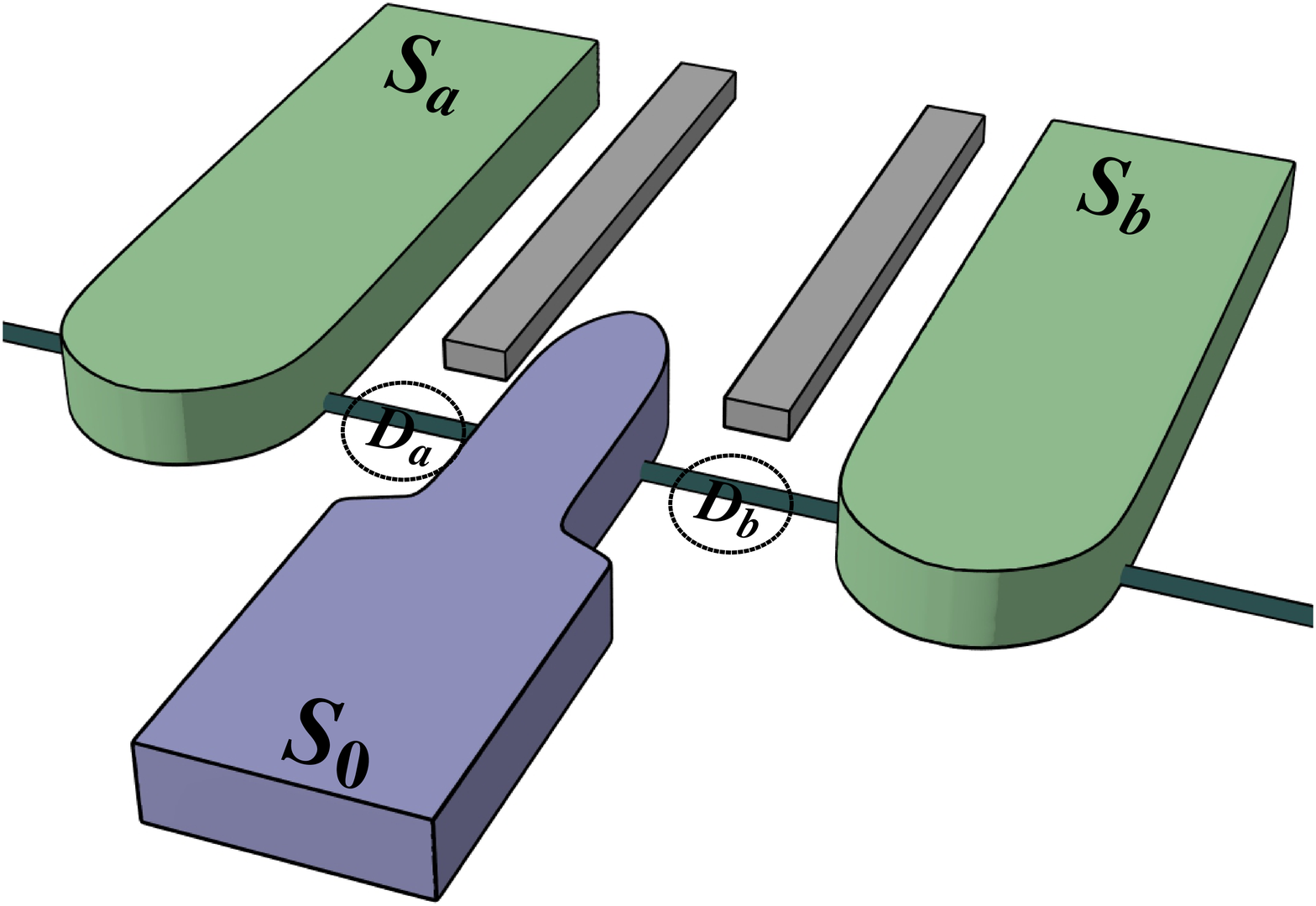}
\includegraphics[width=0.49\columnwidth]{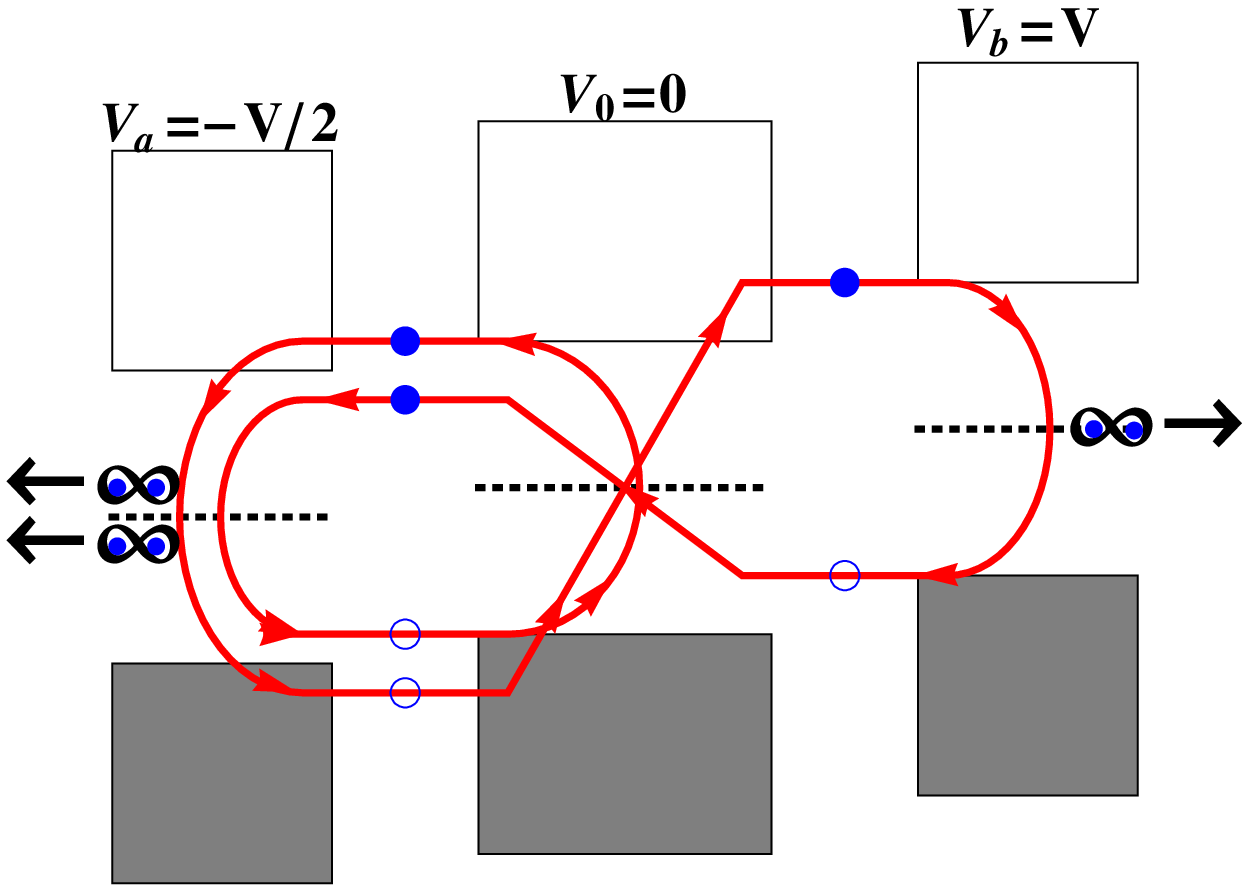}
\caption{(color online) A Josephson bijunction (left). Superconductors $S_{\alpha}$
  ($\alpha = a,b$) are biased at voltages $V_{\alpha}$, while $S_0$ is grounded. The distance between
  the two quantum dot junctions is comparable to the
  coherence length. The right panel shows the energy diagram for the
  $S_aD_aS_0D_bS_b$ bijunction and a higher order diagram associated with a
  ``sextet'' current with 3 pairs emitted, 2 in $S_a$ and 1 in $S_b$, with $V_a=-V_b/2$.}
\label{fig:bijunction}
\end{figure}

A simple phase argument \cite{FREYN} suggests the existence of quartet resonances 
in a bijunction. Starting with an equilibrium situation,
the current-phase relation of a single tunnel junction $S_aS_0$ with phases $\varphi_a$ in $S_a$ and
$\varphi_0$ in $S_0$ is
$I_{c}\sin\left(\varphi_a-\varphi_0\right)$, to which higher-order
harmonics can also contribute. In a $S_aS_0S_b$ bijunction (with phase $\varphi_b$ in $S_b$), 
there exists in addition a quartet and a pair cotunneling supercurrent.
 The DC quartet supercurrent can be viewed as a non-local second-order harmonic
\begin{equation}
I_Q=I_{Q0}\sin(\varphi_a+\varphi_b-2\varphi_0) \; ,
\end{equation}
while the pair cotunneling
corresponds to a DC Josephson effect between $S_a$ and $S_b$ through $S_0$\cite{FREYN}:
\begin{equation}
I_{PC}=I_{PC0}\sin(\varphi_a-\varphi_b) \; .
\end{equation} 

More generally, assuming large enough transparencies, multipair 
currents $I_{a/b}$ in electrodes $S_a$/$S_b$ are obtained when differentiating 
the Josephson free energy with respect to the superconducting phases (assuming $\varphi_0\equiv 0$)
\begin{equation}
I_{a/b} = \sum_{n,m}
I_{a/b,(n,m)} \sin(n \varphi_a + m \varphi_b) \;.
\end{equation}
When voltages $V_{a/b}$ are applied to $S_{a/b}$, $\varphi_a$ and $\varphi_b$ acquire a time dependence, 
and in the special case where 
\begin{equation}
nV_a + mV_b=0 ,
\end{equation} the adiabatic approximation
yields 
\begin{equation}
d\left (n \varphi_a(t) + m \varphi_b(t)
\right)/dt = 0 .
\end{equation} The corresponding current component
\begin{equation}
I_{a/b,(n,m)}\sin( n \varphi_a(t) + m \varphi_b(t) )
\end{equation}and its higher
harmonics are constant in time despite the applied voltages, thus leading to a DC current signaling the existence of a multipair resonance. 
An example of such a resonance is provided in Fig.~\ref{fig:bijunction} 
from a diagrammatic point of view, showing the case $2 V_a + V_b =0$ to lowest order. 
The voltage constraint allows to close a resonance path provided by one Andreev reflection 
in $S_b$ and $S_0$, two in $S_a$ as well as two CAR amplitudes in $S_0$. 
 Note that in general these multipair resonances
must coexist with the usual AC components
\begin{equation}
 I_{a,(1,0)}=I_{0a}\sin\varphi_a(t) \quad \mbox{,} \quad
I_{b,(0,1)}=I_{0b}\sin\varphi_b(t) ,
\end{equation} 
and with the MAR DC currents
discussed in Ref.~[\onlinecite{HOUZET-SAMUELSSON}]. 
While this low-bias argument suggests the possibility of multipair 
resonances also at higher voltages, an exact out-of-equilibrium 
calculation at arbitrary voltage is still lacking, and it is discussed below. 

\section{Hamiltonian formalism}
\label{sec:hamil}
The model Hamiltonian of the $S_a D_a S_0 D_b S_b$ bijunction is written as 
\begin{equation}
\hat{\cal H} =  \sum_{j} \hat{\cal H}_j +\hat{\cal H}_D + \hat{\cal H}_T
\end{equation}
where $\hat{\cal H}_j $ is the Hamiltonian for the lead $S_j$
($j=0,a,b$), expressed with the Nambu spinors
\begin{equation}
\hat{\cal H}_j = \sum_k \Psi^\dagger_{jk} \left(
\xi_k \, \sigma_z + \Delta \, \sigma_x \right) \Psi_{jk} ~,~~~
\Psi_{jk} = \left(
\begin{array}{c}
\psi_{jk, \uparrow} \\
\psi^\dagger_{j(-k), \downarrow}
\end{array} \right),
\end{equation}
with the Pauli matrices acting in the Nambu space.
$\hat{\cal H}_D$ is the Hamiltonian of the two dots, with a single non-interacting level in each dot:
\begin{equation}
\hat{\cal H}_D=\sum_{s,\alpha=a,b} \varepsilon_{\alpha} d^{\dagger}_{\alpha s}d_{\alpha s}.
\end{equation}
$\hat{\cal H}_T$ is for the tunneling between the dots and the electrodes:
\begin{equation}
\hat{\cal H}_T(t) = \sum_{jk\alpha} \,
\Psi^\dagger_{jk} \, t_{j\alpha}   e^{i \sigma_z \varphi_j/2} \, {\bf d}_{\alpha} + {\rm h.c.}  ~,
\label{H_T}
\end{equation}
where ${\bf d}_{\alpha} = (d_{\alpha\uparrow} \; d^{\dagger}_{\alpha\downarrow})$ 
is the Nambu spinor for dot $\alpha$, and $t_{j\alpha}$ is the tunneling amplitude between lead $j$ and dot $\alpha$.

The phases are specified by the applied voltages $\varphi_j(t) = \varphi_{j}^{(0)} + 2e V_jt/\hbar$.  The ``bare'' phases $\varphi_{j}^{(0)}$, which are usually unimportant in an out-of-equilibrium setup,
are relevant here in the transport calculations.  The
superconducting gaps $\Delta$ are assumed identical and the couplings are
taken symmetric. The width of the superconducting region $S_0$ is assumed 
to be negligible, a situation which corresponds to the maximum coupling between 
the two junctions forming the bijunction.

As the leads degrees of freedom are quadratic, they can be integrated out by averaging the evolution operator
 over these leads. We use for this a Keldysh path-integral technique.
 The Green's function $\hat{G}$ of the dots which is non-perturbative in $\hat{\cal H}_T$,
is obtained from a Dyson equation \cite{CUEVAS} involving the free dots Green function, 
and electrode self-energies with both local and non-local propagators.\cite{JONCKHEERE}
The details for a multi-terminal structure with two quantum dots have
 been given in Ref.~[\onlinecite{CHEVALLIER}]. Due to the presence of the two dots, the Green function of the dots is a 2x2 matrix in the dots space:
\begin{equation}
\check{G}_{\alpha\beta}^{\eta \eta'}(t,t')=-i\left\langle T_C\left\{{\bf d}^{\eta}_{\alpha}(t){\bf d}^{\dagger \eta'}_{\beta}(t')\right\}\right\rangle \; ,
\end{equation} 
where $\eta, \eta'$ are Keldysh indices.
 The self-energy is also a 2x2 matrix in the dots space:
 \begin{equation}
 \check{\Sigma} = \left( \begin{array}{cc}
              \hat{\Sigma}_{aa} & \hat{\Sigma}_{ab} \\
              \hat{\Sigma}_{ba} & \hat{\Sigma}_{bb} 
              \end{array} \right) \; ,
 \end{equation}
 and the component $\Sigma_{\alpha \beta}$ (with $\alpha,\beta = a,b$) is given by
 a sum over the leads $j$:
\begin{widetext}
 \begin{align}
 \label{self_energy_onedot}
\hat{\Sigma}_{\alpha \beta}(t_{1},t_{2})=
\sum_{j}
 \Gamma_{j,\alpha \beta}\int_{-\infty}^{\infty}\frac{d\omega}{2\pi} 
      e^{-i\omega(t_{1}-t_{2})} e^{-i\sigma_{z}(V_{j}t_{1}+\varphi_j^{(0)}/2)}   
       [\omega\cdot{1}-\Delta_{j}\cdot\sigma_{x}]e^{+i\sigma_{z}(V_{j}t_{2}+\varphi_j^{(0)}/2)}\\ \notag
   \otimes\left[ -\frac{\Theta(\Delta_{j}-|\omega|)}{\sqrt{\Delta_{j}^{2}-\omega^{2}}} \tau_{z}
    +i\,{\rm sign}(\omega) \frac{\Theta(|\omega|-\Delta_{j})}{\sqrt{\omega^{2}-\Delta_{j}^{2}}}
     \left(\begin{array}{cc}
         2f_{\omega}-1 &~~ -2f_{\omega} \\
        +2f_{-\omega} &~~ 2f_{\omega}-1 \\
\end{array}\right)\right],
\end{align} 
\end{widetext}
where $\Gamma_{j,\alpha \beta} = \pi\nu(0)t^{*}_{j\alpha}t_{j\beta}$ and $f_{\omega}$ is the Fermi function.

The average current from electrode $j$ can then be computed using a Meir-Wingreen 
type formula \cite{Meir Wingreen} generalized to superconductors \cite{JONCKHEERE,CHEVALLIER}:
\begin{align}
\left\langle I_{j\alpha}\right\rangle (t) =&\frac{1}{2}\textrm{Tr}\Bigg\{
  (\tau_z\otimes \sigma_z)\int^{+\infty}_{-\infty}dt'\Big(\check{G}(t,t')\check{\Sigma}_{j}(t',t)- \nonumber \\ 
   & \check{\Sigma}_{j}(t,t')\check{G}(t',t)\Big)_{\alpha\alpha}\Bigg\},
\label{mw}   
\end{align}
where $\tau_z$ acts in Keldysh space, and $\sigma_z$ in Nambu space, and the trace is taken in
the Nambu-Keldysh space. For arbitrary voltages $V_a$ and $V_b$, the time-dependence
of the system is described in terms of two independent Josephson frequencies $\omega_a = 2 e V_a/ \hbar$ 
and $\omega_b=2 e V_b/ \hbar$, the Green function $\hat{G}(t,t^{\prime})$ is a function of two times, 
and solving the Dyson equation is a daunting task. 
However, when the voltages $V_a$ and $V_b$ applied to superconductors $a$ and $b$ are commensurate 
($n V_a + m V_b =0$, with $n$ and $m$ integers), the time-dependence of the system is periodic,
with a period $T = |m| 2\pi/\omega_a = |n| 2\pi /\omega_b$, where $\omega_{a,b} = 2 e|V_{a,b}|/\hbar$ are
the Josephson frequencies.  As in the study of standard 
multiple Andreev reflection (MAR) between two superconductors\cite{JONCKHEERE},
it is then convenient to introduce the double Fourier transforms with summation over discrete domains
in frequency:
\begin{align}
\check{G}(t,t^{\prime})&=&\sum\limits_{n,m=-\infty}^{+\infty}\int_{F}\frac{d\omega}{2\pi}e^{-i\omega_{n}t+i\omega_{m}t^{\prime}}\check{G}_{nm}(\omega)~,
\label{eq:doublefourier1}\\
\check{\Sigma}(t,t^{\prime})&=&\sum\limits_{n,m=-\infty}^{+\infty}\int_{F}\frac{d\omega}{2\pi}e^{-i\omega_{n}t+i\omega_{m}t^{\prime}}\check{\Sigma}_{nm}(\omega)~,
\label{eq:doublefourier2}
\end{align}
where $\omega_{n}=\omega+n \tilde{V}$, the frequency integration is performed over a
finite domain $F\equiv [-\tilde{V}/2,\tilde{V}/2]$, and $\tilde{V}$ is the smallest common 
mutiple of $|V_a|$ and $|V_b|$.
The advantage of this representation is that the Dyson equation for the full Green function $\check{G}_{nm}(\omega)$
is now a matrix equation:
\begin{equation}
\check{G}_{nm}(\omega)=\left[\check{G}_{0,nm}^{-1}(\omega)-\check{\Sigma}_{nm}(\omega)\right]^{-1}~,
\label{eq:dressedG}
\end{equation}
where $\check{G}_{0,nm}$ is the dots Green function without coupling to the superconducting leads. This
equation can be solved by limiting the discrete Fourier transforms to a cutoff energy $E_c$, which gives finite
matrices in Eq.(\ref{eq:dressedG}). The cut-off energy $E_c$  must be chosen large compared to all the relevant energies
in the system. $E_c$ defines a finite number of frequency domains $n_{max}$. 
As the width of each domain is $\sim V$, one has $n_{max} \sim \Delta/V$, which implies that obtaining numerically
the full Green function becomes very expensive at very low voltage. Typical values which we have used in our calculations
are in the range $E_c \sim 5$ to $ 10 \Delta$.

From Eq.(\ref{self_energy_onedot}), we find that the self-energy in the double Fourier representation is
(writing explicitely the 2x2 matrix of Nambu space)
\begin{widetext}
\begin{equation}
\hat{\Sigma}_{\alpha\beta,nm}(\omega_{n})=\sum_j \Gamma_{j}\left(\begin{array}{cc}
  \delta_{n,m}\hat{X}_{j}(\omega_{n}-\sigma_j V/2) & \delta_{n-\sigma_j,m}\hat{Y}_{j}(\omega_{n}-\sigma_j V/2)
                                                                      e^{-i \varphi_j^{(0)}} \\
  \delta_{n+\sigma_j,m}\hat{Y}_{j}(\omega_{n}+\sigma_j V/2) e^{+i \varphi_j^{(0)}}  
            & \delta_{n,m}\hat{X}_{j}(\omega_{n}+\sigma_j V/2) \\
\end{array}\right)~,
\end{equation}
where $\hat{X}$ and $\hat{Y}$ are matrices in the Keldysh space:
\begin{equation}
\hat{X}_{j}(\omega)=\left[-\frac{\Theta(\Delta_{j}-|\omega|)\omega}{\sqrt{\Delta_{j}^{2}-\omega^{2}}}\hat{\tau}_{z}
+i\frac{\Theta(|\omega|-\Delta_{j})|\omega|}{\sqrt{\omega^{2}-\Delta_{j}^{2}}}\left(\begin{array}{cc}
  2f_{\omega}-1 &~~ -2f_{\omega} \\
  +2f_{-\omega} &~~ 2f_{\omega}-1
\end{array}\right)\right],
\end{equation}
and $\hat{Y}_{j}(\omega)=-\Delta_{j}\hat{X}_{j}(\omega)/\omega$.
The expression of the Fourier transform of the current from dot $\alpha$ to lead $j$ is:
\begin{equation}
\langle I_{j\alpha}\rangle(\omega^{\prime})=\!\!\sum\limits_{n,l}\!\!~2\pi\delta\Big(\omega^{\prime}-(n-l)V\Big)
\frac{1}{2}\int_{F}\frac{d\omega}{2\pi}\mbox{Tr}\Big(\sigma_{z}\hat\tau_{z}
\!\!\sum\limits_{m}\!\!\left[\check{G}_{nm}(\omega)\check\Sigma_{j,ml}(\omega)-
\check\Sigma_{j,nm}(\omega)\check{G}_{ml}(\omega)\right]_{\alpha \alpha}\Big)
,\label{eq01}
\end{equation} 
 \end{widetext}
The DC current, which we study in the following sections,
 is obtained by taking $\omega^{\prime} = 0$ in the last equation. 

\section{Metallic junction regime}
\label{sec:metallic}
\begin{figure*}[tb]
\includegraphics[width=13.cm]{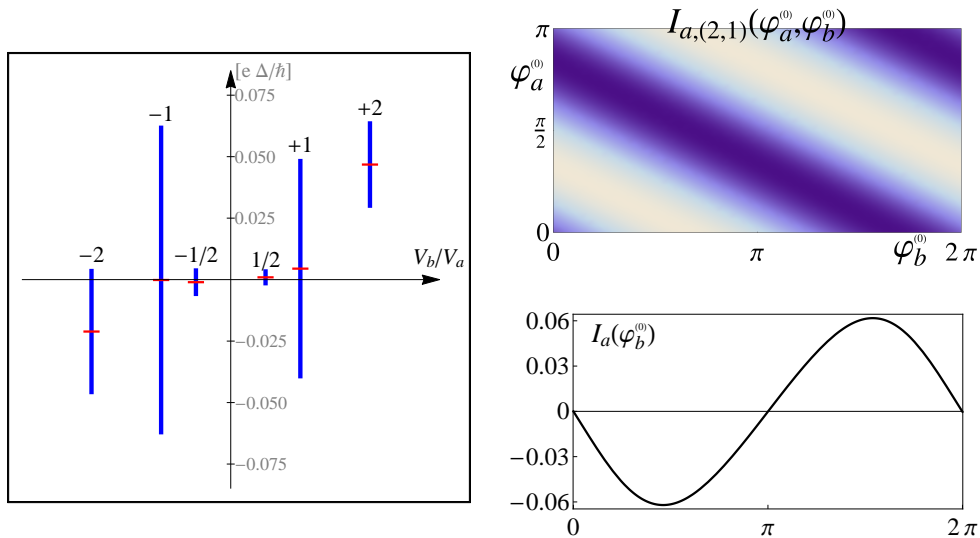}
\caption{(color online) ``Broad'' dots regime (metallic junctions):  $|\varepsilon_{a,b}|=6\Delta$, $\Gamma_{a,b}=4\Delta$. 
    Left: Amplitudes of the phase-dependent DC current $\langle I_a(\varphi_b^{(0)}) \rangle$ for the main resonances (with $|n|+|m| \leq 3$), 
    in units of $e \Delta/\hbar$,
    centered around the values of the phase-independent current (small horizontal bars).
   Horizontal axis is $V_a/V_b$, with $V_b/\Delta=0.3$.
   Upper right: Current $\langle I_a \rangle$, for the resonance $2 V_a + V_b =0$, as a function of the 
   phases $\varphi_a^{(0)}$ and $\varphi_b^{(0)}$, showing the dependence in $2 \varphi_{a}^{(0)} + \varphi_{b}^{(0)}$.
   Lower right: the current-phase relation $\langle I_a(\varphi_b^{(0)}) \rangle$ at $\varphi_a^{(0)}=0$ for the resonance $V_a+V_b=0$,
   which shows the $\pi$-phase behavior.
\label{fig:resonances}}
\end{figure*}

We first consider the regime in which each dot mimics a metallic junction, achieved by placing energy levels out of resonance $\epsilon_{\alpha} > \Delta$ and choosing large couplings $\Gamma_{\alpha}>\Delta$ ($\alpha=a,b$, and
$\Gamma_{\alpha} = \sum_j \pi\nu(0)|t_{j\alpha}|^2$, where $t_{j\alpha}$ are tunneling couplings defined in Eq.~(\ref{H_T}),
and $\nu(0)$ the normal density of states of the electrodes at the Fermi energy) .  
We compute the DC currents $\langle I_{a/b} \rangle$
 for different ratio of the voltages, satisfying
$nV_a+mV_b=0$. 
The results for the largest resonances ($|n|+|m|\leq 3$ in  $nV_a+mV_b=0$) are shown in the left
panel of Fig.~\ref{fig:resonances}.
One clearly sees that the resonances are easily distinguished from the phase-independent background current.   
The resonant multipair DC-current $\langle I_a^{MP}\rangle$ is a function of the combination 
$n \varphi_{a}^{(0)} + m \varphi_{b}^{(0)}$, which implies a simultaneous
crossing of $n$ pairs from $S_0$ to $S_a$ and $m$ pairs from $S_0$ to $S_b$.
The upper right panel in Fig.~\ref{fig:resonances}
shows as an example the phase dependence for $n=2$ and $m=1$,
which is indeed a sinusoidal function of the combination $2\varphi_{a}^{(0)}+\varphi_{b}^{(0)}$. 
The existence of DC phase-coherent resonances despite large nonzero voltages 
is the result of new coherent modes connecting the three superconductors.

One of the lowest-order (and larger) resonances corresponds to quartets ($V_a=-V_b$), i.e. to the correlated transmission of 
two pairs from $S_0$ to $S_a$ and $S_b$ respectively. The ``dual'' lowest-order resonance
corresponds to $V_b=V_a$, where pairs cross from $S_a$ to $S_b$ by
cotunneling through $S_0$.
 The sign of the multipair resonances is non-trivial. In particular, the quartet resonance
is negative, which means that the current $\langle I_a(\varphi_a^{(0)} + \varphi_b^{(0)}) \rangle$ is of $\pi$-type, 
as shown in the lower right panel of Fig.\ref{fig:resonances}.
Similar sign changes of the multipair current-phase relation are also
obtained for certain high-order resonances. 
The $\pi$-shift is understood from a simple argument. It is related to the internal structure of a Cooper pair
via the antisymmetry of its wavefunction, similarly to the $\pi$-junction
behavior of a magnetic junction formed by a quantum dot with a localized spin
\cite{GLAZMAN}. Starting from two Cooper pairs in 
 $S_0$, the production of a non-local quartet consists in forming two non-locally 
 entangled singlets in the dots $D_a$ and $D_b$. 
 These two split pairs correspond to two CAR amplitudes, as those apparent 
 in Figure \ref{fig:bijunction}. 
 A non-local singlet is obtained by the operator 
$ \frac{1}{\sqrt{2}}(d^{\dagger}_{a \uparrow} d^{\dagger}_{b \downarrow} - 
 d^{\dagger}_{a \downarrow} d^{\dagger}_{b \uparrow} )$ acting on the empty dots.
 Applying this operator twice to describe a non-local quartet state leads to
$
\Psi_{Q,D_a,D_b} = 
-|\uparrow \downarrow\rangle_{a}\, |\downarrow \uparrow\rangle_{b} ,
$
which is recast as the opposite of the product of a pair in $D_a$ and another
one in $D_b$. A similar reasoning can be applied in order to explain the anomalous 
sign of higher-order harmonics.

\section{Resonant dots regime}
\label{sec:resonant}
We now investigate the possibility for optimizing the multipair resonances by 
tuning the dot levels, with  
$\varepsilon_a = - \varepsilon_b = - 0.4\Delta$ inside the gap, choosing
small values of the couplings $\Gamma_a = \Gamma_b=0.1 \Delta$.
We focus on the quartet resonance $V_a=-V_b$ for specificity (similar behavior
is observed for the other resonances).

\begin{figure}[tb]
\centerline{\includegraphics[width=0.9\columnwidth]{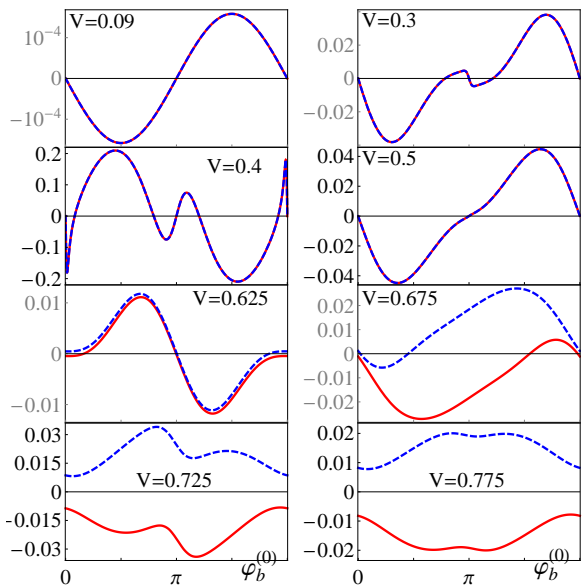}}
\caption{(color online) Current-phase relations $I_a(\varphi_b^{(0)})$ (red, full curve)
 and $I_b(\varphi_b^{(0)})$ (blue, dashed curve), in units of $e \Delta/\hbar$, in the quartet configuration 
$V_a = - V_b$, for the resonant dots regime: $\varepsilon_a=-\varepsilon_b=0.4 \Delta$ and $\Gamma=0.1 \Delta$.
 Note the different $y$ scales in the different panels.
\label{fig:criticalcurrent}}
\end{figure}

When the bias is small enough ($V_b \alt 0.1 \Delta$ here),
the system is in the adiabatic regime, and the current does not change  
when $V_b$ is varied.
We independently checked with a Matsubara formalism calculation (not shown) 
that this current is the same as the one obtained here 
at equilibrium ($V_b=0$). The current-phase relations $\langle I_a(\varphi_b^{(0)})\rangle$ and 
$\langle I_b(\varphi_b^{(0)})\rangle$ for $V=0.09 \Delta$ are shown  
in the first panel of Fig.~\ref{fig:criticalcurrent}. 
These average currents are identical, thus are made only from a quartet component. They
show a purely harmonic function of the phase $\varphi_b^{(0)}$,
and suggest a $\pi$-junction behavior for the quartet resonance near equilibrium.

\begin{figure}
\includegraphics[width=4.cm]{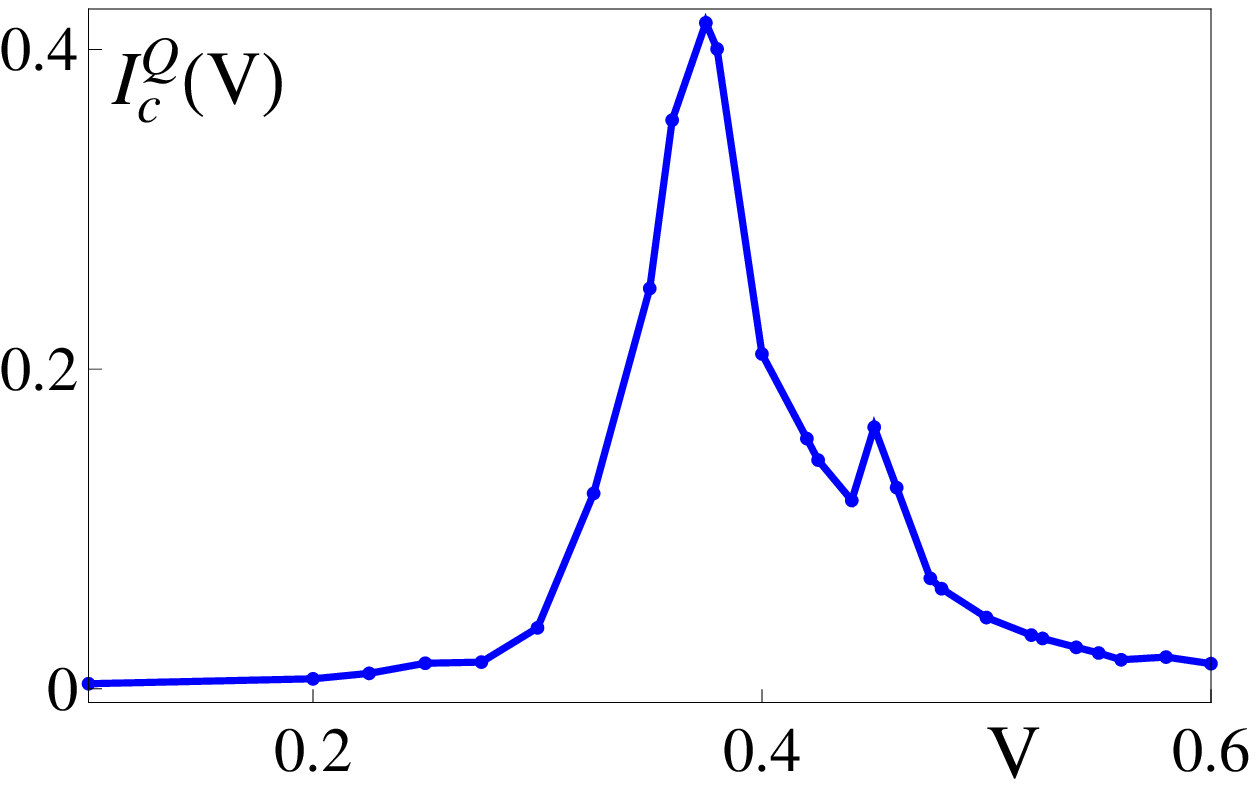}
\hspace{.7cm}
\includegraphics[width=3.5cm]{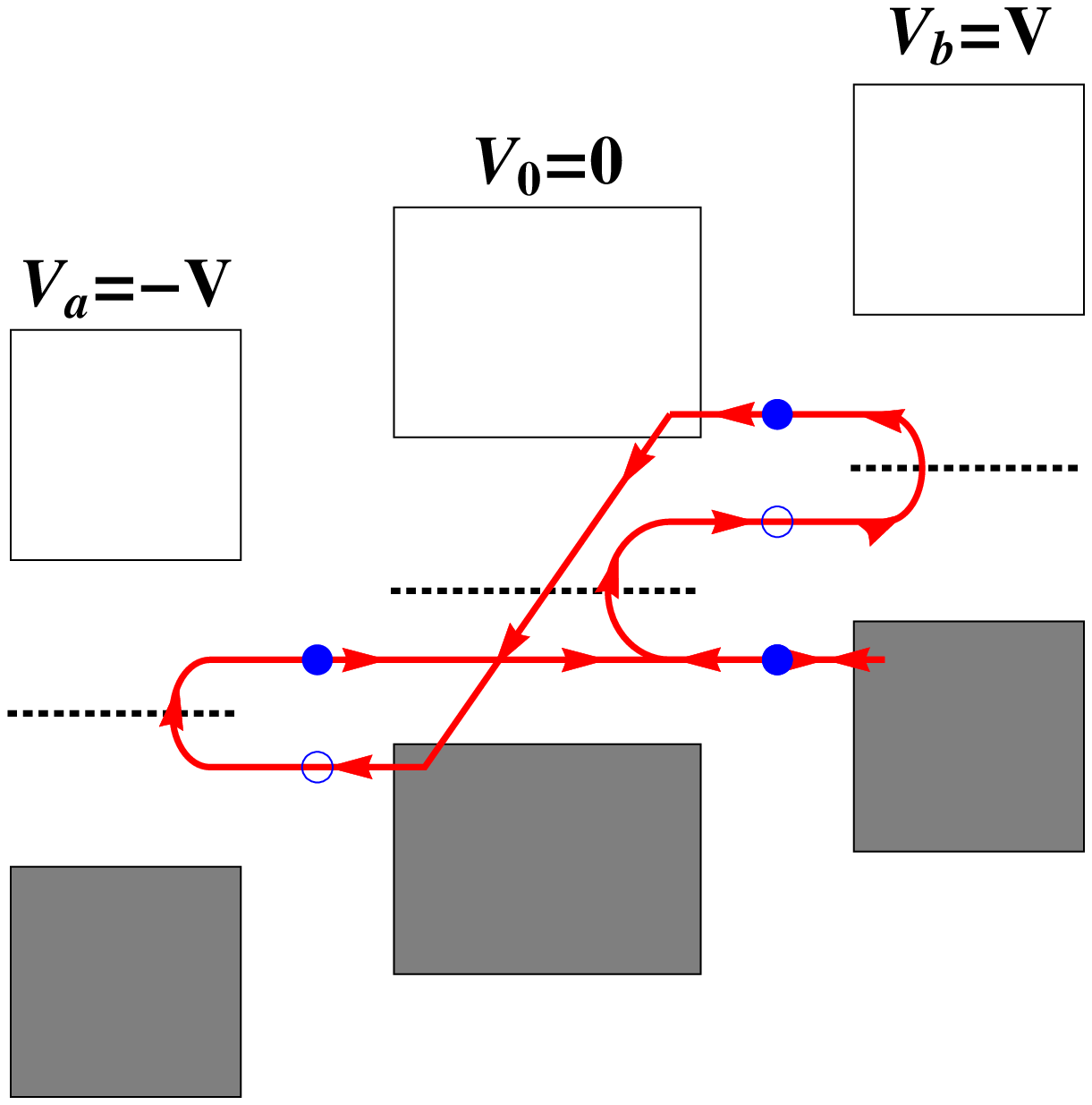}
\caption{(color online)
	Left panel: critical current $I_c^{Q}$ in units of $e \Delta/\hbar$ 
	as a function of the voltage in the quartet configuration, for resonant dots
	 (same parameters as in Fig.~\ref{fig:criticalcurrent}). 
	 Right panel: Lowest order diagram contributing to the phase-dependent MAR current.}
\label{fig:ProcMixtes}
\end{figure}

When $V_b$ increases and the non-adiabatic regime is reached,
drastic changes appear in the current-phase relations, as shown in the next panels
of Fig.~\ref{fig:criticalcurrent}. There, both the sign and the (non-sinusoidal) shape 
of the current-phase relation changes rapidly with $V_b$ as it approaches 
the dot energy $|\varepsilon_b|$.
The amplitude of the quartet current near the resonance is $\sim 1000$ times larger than the one in the
adiabatic regime. This resonant effect of the dot levels is most apparent by plotting the critical current $I_c^{Q}$
(the maximum of the absolute value of the phase-dependent part of $I_a(\varphi_b^{(0)})$)
as a function of $V_b$. This is shown in the left panel of Fig.~\ref{fig:ProcMixtes}. 
$I_c^{Q}$ sharply increases and reaches a maximum around $V_b\simeq \varepsilon_b$.
The large increase in the quartet current is due to a double resonant effect:
first, as the dots have opposite energies $\varepsilon_a = -\varepsilon_b$,
the formation of a quartet in the double dot as a pair in $D_a$
and a pair in $D_b$ is resonant (this is true for any voltage $V_b$); second 
when $V_b\simeq \varepsilon_b$ the tunneling of a pair from $D_a$ to $S_a$, and from $D_b$ to $S_b$, is also resonant.

Increasing $V_b$ further, e.g.  $V_b \agt 2\Delta/3$ in the present case, we see from the lower panels of Fig.~\ref{fig:criticalcurrent} that the currents 
$\langle I_a(\varphi_b^{(0)})\rangle$ and $\langle I_b(\varphi_b^{(0)})\rangle$
start to deviate substantially.  
This implies the existence of another phase-sensitive process different from the one responsible for multipair resonances.
  We call this current contribution 
$I^{phMAR}$ (for phase-sensitive MAR), as it is the result of the combination of a multipair  process with MAR. 
The lowest order diagram contributing to $I^{phMAR}$ is shown in the right panel of Fig.~\ref{fig:ProcMixtes}.
It can be seen as the interference of the amplitudes of two MAR processes at the $S_aS_0$ and $S_bS_0$ interfaces,
each promoting a quasiparticle from an energy $\sim -\Delta$ in superconductor $S_b$ to an energy $\sim +\Delta$ in superconductor $S_0$. This diagram has a threshold at $V=2 \Delta/3$, corresponding to the observed value
at which $I^{phMAR}$ becomes noticeable.
However, unlike the usual MAR processes found in single junctions,
 this process (and similar ones of higher order) has the striking property of being phase-dependent.

Addressing more general resonances, the total DC current can be decomposed into $3$ components, as inspired by 
Josephson's work \cite{JOSEPHSON}. 
Defining $\bar I=(\langle I_a\rangle,\langle I_b\rangle)$, $\bar \varphi=(\varphi_a,\varphi_b)$, $\bar V=(V_a,V_b)$,
 $\bar \epsilon=(\epsilon_a,\epsilon_b)$ one has 
$\bar I(\bar \varphi,\bar V, \bar \epsilon)=
 \bar I^{MP}(\bar \varphi,\bar V, \bar \epsilon)+\bar I^{phMAR}(\bar \varphi,\bar V,\bar \epsilon)+ \bar I^{qp}(\bar V,\bar \epsilon)$. 
 Assuming electron-hole symmetry to hold, for instance with flat normal metal density of states in the leads, one can show that in the situation studied here,
the DC-current obeys the relation: 
\begin{equation}
\bar I(\bar\varphi^{(0)},\bar V,\bar \epsilon)=-\bar I(-\bar \varphi^{(0)},-\bar V,-\bar \epsilon) .
\end{equation}
Then the following properties hold:
 (i) The pure quasiparticle current
$\bar I^{qp}$ is phase-insensitive and odd in voltages. 
(ii) The coherent multipair current $\bar I^{MP}$ is a function of
$n \varphi_{a}^{(0)} + m \varphi_{b}^{(0)}$, it is {\it odd} in phases and {\it even} in voltages,
 just like the non-dissipative Josephson term. It satisfies $m \langle I_a\rangle =  n  \langle I_b\rangle$.
 (iii) The component $\bar I^{phMAR}$ is {\it even} in phases and
{\it odd} in voltages, like the dissipative ("$\cos\varphi$") Josephson component, {\it but} it becomes DC in a bijunction. This $\bar I^{phMAR}$ component is also a function of
$n \varphi_{a}^{(0)} + m \varphi_{b}^{(0)}$. 

\section{Conclusions}
\label{sec:conclusions}
We have shown by non-perturbative out-of-equilibrium calculations that coherent multipair and phase-dependent MAR processes appear in a superconducting bijunction. These are due to crossed Andreev reflection processes, through the formation of several entangled non-local pairs, and lead to signatures in the DC current with very specific phase and voltage dependence.
A natural extension of the present work should focus on the role of local Coulomb interaction on the dots. 
In the metallic junction regime and in the resonant regime near the dot resonance, 
a self-consistent mean-field treatment could be applied (as done in Ref.~[\onlinecite{Rech}] for a three terminal
normal-superconducting setup with resonant dots). We expect that the same physical mechanisms would qualitatively produce the same effects.
A more complex treatment would be required away from resonance, where interactions would have a larger impact and
the Kondo mechanism could play an important role.

From an experimental standpoint, multipair resonances can be directly detected by transport measurements where
one probes the nonlocal conductance $d\langle I_a\rangle / dV_b$ as a function of
$V_a$, $V_b$. 
The phase coherence of the multipair current, and its actual dependence in $\varphi_{a/b}^{(0)}$, however, are more difficult to probe directly.  One way would be to design specific SQUID geometries or microwave reflectivity experiments.\cite{REFLECTIVITY}

\smallskip
The authors acknowledge support from ANR contract ``Nanoquartet'' 12-BS-10-007-04, and the ``m\'esocentre'' of 
Aix-Marseille Universit\'e for numerical resources.

\end{document}